\documentclass[aps, prb,showpacs,twocolumn, floatfix]{revtex4}
\usepackage{amsmath}
\usepackage{amssymb}
\usepackage{mathtools}
\usepackage{epsfig}
\usepackage{graphics}
\usepackage{color}
\usepackage{bm}
\usepackage{amssymb}
\usepackage{graphicx}
\begin{document}

\title{Reply to Comment on "Does the Equivalence between Gravitational Mass and Energy Survive for a Composite Quantum Body?"}

\author{{\bf Andrei G. Lebed}$^*$}

\affiliation{Department of Physics, University of Arizona, 1118 E.
4th Street, Tucson, AZ 85721, USA}

\author{Correspondence should be addressed to A.G. Lebed:
lebed@physics.arizona.edu}

\begin{abstract}
We have recently shown that both passive and active gravitational
masses of a composite body are not equivalent to its energy due to
some quantum effects. We have also suggested an idealized and more
realistic experiments to detect the above mentioned inequivalence
for a passive gravitational mass. The suggested idealized effect
is as follows. A spacecraft moves protons of a macroscopic
ensemble of hydrogen atoms with constant velocity in the Earth's
gravitational field. Due to non-homogeneous squeezing of space by
the field, electron ground state wave function experiences
time-dependent perturbation in each hydrogen atom. This
perturbation results in the appearance of a finite probability for
an electron to be excited at higher energy levels and to emit a
photon. The experimental task is to detect such photons from the
ensemble of the atoms. More realistic variants of such experiment
can be realized in solid crystals and nuclei, as first mentioned
by us. In his recent Comment on our paper, Crowell has argued that
the effect, suggested by us, contradicts the existing experiments
and, in particular, astronomic data. We show here that this
conclusion is incorrect and based on the so-called "free fall"
experiments, where our effect does not have to be observed.
\end{abstract}

\maketitle

\section{Introduction}
Creation of the so-called Theory of Everything is well known to be
one of the most important problems in physics. It is also known
that development of the Quantum Gravitation theory is one of the
most important steps in this direction. Nevertheless, the latter
problem appears to be extremely difficult. One of the reasons for
that is the fact that the foundations of  General Relativity and
Quantum Mechanics are very different. Another reason is the
absence of the corresponding experimental data. We recall that, so
far, quantum effects have been directly tested only in the
Newtonian variant of gravitation (see, for example, Refs. [1,2]).
In this complex situation, we have recently suggested two novel
phenomena [3-9]. In particular, we have demonstrated that both
passive and active gravitational masses of a composite body are
not equivalent to its energy due to some quantum effects. We have
also suggested two experimental ways [3-9] to test the above
mentioned phenomena. If one of such experiments is done, it will be
the first direct observation of quantum effects in General
Relativity.

\section{Goal}

Very recently, Crowell has published a Comment [10] on our paper
[5], which criticizes one of the suggested by us experiments,
which can demonstrate inequivalence of passive gravitational mass
of a composite quantum body and its energy. The idealized variant
of the experiment is as follows. There is a macroscopic ensemble
of hydrogen atoms with each of them being in ground state at
$t=0$. Protons of all atoms are dragged by a spacecraft with
constant velocity in the Earth's gravitational field. Due to
non-homogeneous squeezing of space by the gravitational field, the
atoms are shown [3-7,9] become excited and emit photons. As
mentioned in Ref.[3], the above described phenomenon is very
general and have to be observed in solids, nuclei, and elementary
particles. The main criticism of the experiment [3-7,9] in Comment
[10] is the statement that the application of our theory to
experiments on proton decay is not consistent with the existing
experimental data. The goal of our Reply is three-fold. First, we
pay attention that the discussed in Ref. [10] existing
experimental data are obtained for free falling objects. On the
other hand, the idealized experiment, suggested by us [3-7,9],
corresponds to transportation of centers of masses of the hydrogen
atoms (i.e., protons) by spacecraft with a constant velocity. We
stress that these are two different types of experiments. Second,
to strengthen our arguments, we derive the Hamiltonian for the
transportation of a hydrogen atom with constant velocity,
semi-quantitatively introduced in Refs. [3-7,9], from the Dirac
equation in a curved spacetime of General Relativity. Third, we
discuss "free fall" experiments for a hydrogen atom and make the
conclusion that the effect, suggested by us for passive
gravitational mass [3-7,9], does not have to be observed under
such conditions. Thus, proton decay does not have to demonstrate
our effect in "free fall" experiments too. So, we make a
conclusion that, contrary to the statement of Comment [10], the
existing experiments on proton decay do not contradict to our
theoretical results.

\section{Semi-Quantitative Hamiltonian}
First, let us derive the Hamiltonian of Refs.[3-7,9] for a hydrogen
atom in the Earth's gravitational field, using semi-quantitative
approach. Below, we consider the case of a weak gravitational
field, therefore, we can write the standard interval, describing
spacetime in a weak field approximation [11]:
\begin{eqnarray}
&ds^2 = - \biggl( 1 + 2\frac{\phi}{c^2} \biggl)(c dt)^2 + \biggl(
1 - 2 \frac{\phi}{c^2} \biggl) (dx^2 +dy^2+dz^2 ), \nonumber\\
&\phi = - \frac{GM}{R}.
\end{eqnarray}
[Here $G$ is the gravitational constant, $c$ is the velocity of
light, $M$ is the Earth mass, and $R$ is the distance between its
center and proton.] In accordance with General Relativity, we
introduce the so-called local proper spacetime coordinates,
\begin{eqnarray}
&x'=\biggl(1-\frac{\phi}{c^2} \biggl) x, \ \ \ y'=
\biggl(1-\frac{\phi}{c^2} \biggl) y,
\nonumber\\
&z'=\biggl(1-\frac{\phi}{c^2} \biggl) z , \ \ \ t'=
\biggl(1+\frac{\phi}{c^2} \biggl) t,
\end{eqnarray}
where space coordinates do not depend on time and where the
interval (1) has the Minkowski form.

In these local spacetime coordinates, we can approximately write
the Schr\"{o}dinger equation for electron in the atom in the
standard form,
\begin{equation}
i \hbar \frac{\partial \Psi({\bf r'},t')}{\partial t'} = \hat H_0
(\hat {\bf p'},{\bf r'}) \ \Psi({\bf r'},t')  ,
\end{equation}
\begin{equation}
\hat H_0 (\hat {\bf p'},{\bf r'})= m_ec^2 - \frac{\hat {\bf
p'}^2}{2 m_e} - \frac{e^2}{r'},
\end{equation}
where proton is supposed to have a fixed position due to action of
some non-gravitational force on it. [Here $\hat{\bf p'} = -i \hbar
\partial/\partial {\bf r'}$; $m_e$ and $e$ are electron mass and
charge, respectively.] Let us discuss the approximation (1)-(4).
First, in Eqs.(1),(2), we take into account only terms of the
order of $|\phi|/c^2$, which can be estimated as $10^{-9}$ near
the Earth. Second, in Eqs.(3),(4), we disregard the so-called
tidal effects. This means that we do not differentiate
gravitational potential, $\phi$, with respect to electron
coordinates, ${\bf r}$ and ${\bf r'}$. In the next section, we
estimate the tidal terms in the Hamiltonian, which, as will be
shown, are of the order of $(r_B/R_0)|\phi/c^2| (e^2/r_B) \sim
10^{-17}|\phi/c^2| (e^2/r_B)$ in the Earth's gravitational field.
[Here $r_B$ is a hydrogen atom typical "size" (i.e., the Bohr's
radius), $R_0$ is the Earth's radius.] Third, we consider proton
as a classical particle with mass $m_p \gg m_e$, whose position is
fixed and kinetic energy is negligible. As usual, we treat the
weak gravitation (1),(2), as a perturbation in the inertial
coordinate system, corresponding to the coordinates $(x,y,z,t)$ in
Eq.(2). By substituting of these coordinates in the Hamiltonian
(3),(4), it is easy to obtain the following effective electron
Hamiltonian:
\begin{equation}
\hat H(\hat {\bf p},{\bf r}) = m_e c^2 + \frac{\hat {\bf
p}^2}{2m_e}-\frac{e^2}{r} + m_e  \phi + \biggl( 3 \frac{\hat {\bf
p}^2}{2 m_e} -2\frac{e^2}{r} \biggl) \frac{\phi}{c^2}
\end{equation}
and to rewrite it in more convenient form:
\begin{equation}
\hat H(\hat {\bf p},{\bf r}) = m_e c^2 + \frac{\hat {\bf
p}^2}{2m_e} -\frac{e^2}{r} + \hat m_g (\hat {\bf p},{\bf r}) \phi
\ .
\end{equation}
We point out that, in Eq.(6), we introduce the following expression for
electron passive gravitational mass operator:
\begin{equation}
\hat m_g (\hat {\bf p},{\bf r})  = m_e  + \biggl(\frac{\hat {\bf
p}^2}{2m_e} -\frac{e^2}{r}\biggl) \frac{1}{c^2} + \biggl(2
\frac{\hat {\bf p}^2}{2m_e}-\frac{e^2}{r} \biggl) \frac{1}{c^2} ,
\end{equation}
which is equal to electron weight operator in the weak
gravitational field (1). Note that, in Eq.(7), the first term is
the bare electron mass, $m_e$, the second term corresponds to the
expected electron energy contribution to the mass operator,
whereas the third term is the non-trivial virial contribution to
the gravitational mass operator. We recall that the Hamiltonian
(6),(7) is derived for the case, where a hydrogen atom center of
mass (i.e., proton) has a fixed position with respect to the
Earth. In other words, it is supported in the gravitational field
(1) by some non-gravitational force. Now, suppose that the proton
is dragged with small and constant (with respect to the Earth)
velocity, $u \ll \alpha c$, by a spacecraft, where $\alpha$ is the
fine structure constant and $\alpha c$ is a characteristic value
of electron velocity in a hydrogen atom. In this case, we can use
adiabatic approximation [3-7,9], which results in the following
perturbation for the electron Schr\"{o}dinger equation:
\begin{equation}
\hat V (\hat {\bf p},{\bf r}, R, t)  = + \biggl(2 \frac{\hat {\bf
p}^2}{2m_e}-\frac{e^2}{r} \biggl) \frac{\phi(R+ut)}{c^2}.
\end{equation}
Note that we are interested in electron excitations, therefore, in
the electron Hamiltonian (8), we keep only the virial term, which
does not commute with the Hamiltonian, taken in the absence of
gravitational field. Since the Hamiltonian (8) is time dependent
it cause to the appearance of electron excitations and, thus, to
the appearance of photon emission from a macroscopic ensemble of
the atoms. It is important that the Hamiltonians (6)-(8) are not
valid for the free falling atoms, where we have to introduce the
so-called normal Fermi coordinates [13,14]. As a result free
falling atoms "feel" only second derivatives of the gravitational
potential [13,14].

\section{The Most General Hamiltonian}
To strengthen our arguments, in this section, we derive our
Hamiltonian (6),(7) from the more general Hamiltonian of Ref.[12].
It is obtained from Dirac equation in curved spacetime of General
Relativity. In Ref.[12], completely different physical effect -
the mixing effect between even and odd wave functions in a
hydrogen atom (i.e., the so-called relativistic Stark effect) - is
studied. It is important that it is studied not for the free
falling atoms but for the atom, whose center of mass is supported
by non-gravitational force in the weak gravitational field (1).
Note that the corresponding Hamiltonian is derived in $1/c^2$
approximation, as in our case. The peculiarity of the calculations
of Ref.[12] is that not only terms of the order of $\phi/c^2$ are
calculated, as in our case, but also terms of the order of
$\phi'/c^2$, where $\phi'$ is a symbolic derivative of $\phi$ with
respect to relative electron coordinates in the atom. Note that,
in accordance with the existing tradition, we call the latter
terms tidal ones. Obtained in Ref.[12] the Hamiltonian (3.24) for
the corresponding Schr\"{o}dinger equation can be expressed as a
sum of the following four terms:
\begin{equation}
\hat H (\hat {\bf P}, \hat {\bf p}, \tilde {\bf R},r)= \hat H_0 +
\hat H_1 + \hat H_2 + \hat H_3 \ ,
\end{equation}
where
\begin{equation}
\hat H_0  = m_e c^2 + m_p c^2 + \biggl[\frac{\hat {\bf P}^2}{2(m_e
+ m_p)} + \frac{\hat {\bf p}^2}{2 \mu} \biggl] - \frac{e^2}{r} ,
\end{equation}

\begin{eqnarray}
&\hat H_1  = \biggl\{ m_e c^2 + m_p c^2 + \biggl[3 \frac{\hat {\bf
P}^2}{2(m_e + m_p)} + 3 \frac{\hat {\bf p}^2}{2 \mu} - 2
\frac{e^2}{r} \biggl]\biggl\}
\nonumber\\
&\times \biggl( \frac{\phi - {\bf g}\tilde
{\bf R}}{c^2} \biggl),
\end{eqnarray}

\begin{eqnarray}
&\hat H_2  = \frac{1}{c^2} \biggl(\frac{1}{m_e}-\frac{1}{m_p}
\biggl)[-({\bf g}{\bf r}) \hat {\bf p}^2 + i \hbar {\bf g} \hat
{\bf p}]
\nonumber\\
&+\frac{1}{c^2} {\bf g} \biggl(\frac{\hat {\bf s_e}}{m_e} -
\frac{\hat {\bf s_p}}{m_p} \biggl) \times \hat {\bf p} + \frac{e^2
(m_p-m_e)}{2(m_e+m_p)c^2} \frac{{\bf g}{\bf r}}{r},
\end{eqnarray}

\begin{eqnarray}
&\hat H_3  = \frac{3}{2}\frac{i \hbar {\bf g}{\bf
P}}{(m_e+m_p)c^2} +\frac{3}{2} \frac{{\bf g}{\bf(s_e+s_p)}\times
{\bf P}}{(m_e+m_p)c^2}
\nonumber\\
&- \frac{({\bf g}{\bf r})({\bf
P}{\bf p})+({\bf P}{\bf r})({\bf g}{\bf p})-i\hbar {\bf g}{\bf
P}}{(m_e+m_p)c^2},
\end{eqnarray}
[Here, ${\bf g}=-G \frac{M}{R^3} {\bf R}$]. Let us describe
notations in Eqs.(9)-(13). Note that $\tilde {\bf R}$ and ${\bf
P}$ are position and momentum of a center of mass of the atom,
correspondingly. On the other hand, ${\bf r}$ and ${\bf p}$ are
relative electron position and momentum in the center of mass
coordinate system; $\mu = m_e m_p /(m_e + m_p)$ is the reduced
electron mass. As seen from Eq.(10), $\hat H_0 (\hat {\bf P}, \hat
{\bf p}, r)$ corresponds to the Hamiltonian of a hydrogen atom in
the absence of the external field. We point out that $\hat H_1
(\hat {\bf P}, \hat {\bf p}, \tilde {\bf R}, r)$ corresponds to
couplings of the bare electron and proton masses as well as
electron kinetic and potential energies with the gravitational
field (1). The Hamiltonians $\hat H_2 (\hat {\bf p}, {\bf r})$ and
$\hat H_3 (\hat {\bf P}, \hat {\bf p}, \tilde {\bf R}, r)$
describe the tidal effects.

Note that, in the previous section, we have semi-quantitatively
derived the Hamiltonian (6),(7). Below, we strictly derive it from
the more general Hamiltonian (9)-(13). First, we use the
approximation, where $m_p/m_e \gg 1$, and, thus, we have $\mu =
m_e$. This allows us to consider proton as a heavy classical
particle. We can fix its position, $\tilde {\bf R}$ = const, in
coordinate system, corresponding to the source of the
gravitational field (1), by putting ${\bf P}= 0$ in the
Hamiltonian (9)-(13). Therefore, we can disregard center of mass
momentum and center of mass kinetic energy. Moreover, as seen from
Eq.(13), $\hat H_3 (\hat {\bf P}, \hat {\bf p}, \tilde {\bf R},
r)=0$ in this case. Moreover, let us estimate the first tidal term
(12) in the Hamiltonian. We recall that $|{\bf g}| \simeq
|\phi|/R_0$. It is important that, $|{\bf r}| \sim \hbar / |{\bf
p}| \sim r_B$ and ${\bf p}^2/(2m_e) \sim e^2/r_B$ in a hydrogen
atom. These allow us to evaluate the Hamiltonian (12) as $H_2 \sim
(r_B/R_0) (\phi/c^2) (e^2/r_B) \sim 10^{-17} (\phi/c^2)
(e^2/r_B)$, which is $10^{-17}$ smaller than $H_1 \sim (\phi/c^2)
(e^2/r_B)$. Therefore, we can also disregard the first tidal term
(12) in the total Hamiltonian (9)-(13). As a result, the
Hamiltonian (9)-(13) can be rewritten in a familiar way:
\begin{equation}
\hat H (\hat {\bf p},r)= \hat H_0 (\hat {\bf p}, r) + \hat H_1
(\hat {\bf p}, r)
\end{equation}
\begin{equation}
\hat H_0 (\hat {\bf p}, r) = m_e c^2 + \frac{\hat {\bf p}^2}{2m_e}
- \frac{e^2}{r} ,
\end{equation}
\begin{equation}
\hat H_1 (\hat {\bf p}, r) =  \biggl\{ m_e c^2 + \biggl[3
\frac{\hat {\bf p}^2}{2 m_e} - 2 \frac{e^2}{r} \biggl]\biggl\}
\biggl( \frac{\phi}{c^2} \biggl),
\end{equation}
where we place the proton at the point $\tilde R = R$. Thus, we
can make a conclusion  that the Hamiltonian (14)-(16), derived in
this section, exactly coincides with the Hamiltonian,
semi-quantitatively derived by us earlier [3-7,9] [see
Eqs.(6),(7)].

\section{What is Right and What is Wrong?}
As earlier as in Ref.[3], we concluded that the suggested by us
effect was very general. In particular, we proposed [3] to use it
not only in atomic physics, but also in condensed matter physics
[3,9], nuclear physics [3,10], and elementary particle physics
[3,10]. Here, we recall the physical meaning of the effect. Some
quantum macroscopic system is placed in spacecraft and dragged
with small constant velocity in an external weak gravitational
field. In this case, due to nonhomogeneous squeezing of space by
the field, there appear some quantum excitations in the system,
which result in emission of photons [3-7], phonons [9], pions [10]
or some other particles. The experimental task is to detect these
particles. We pay attention that, in all our previous works
[3-7,9] as well as in the previous sections of the current paper,
we consider the case, where center of mass of a composite quantum
system is dragged by spacecraft. It is important that it is
dragged by means of non-gravitational forces with constant
velocity with respect to source of gravity. We claim that the
extension of our effect to free falling bodies, performed in the
Comment [10], is not legitimate. It is clear seen from papers
[13,14], where examples of a free falling hydrogen atom is
considered and the Fermi normal coordinates are used. As stressed
in Refs.[13,14], the free falling atoms "feel" only second
derivative of the metric (1) and, thus, cannot exhibit our effect.
This is also true for nuclear versions of free falling experiment,
considered in the Comment [10]. To summarize our effect does not
have to be observed in "free fall" experiments, discussed in [10].
Therefore, the central statement of Comment [10], that considered
there nucleus experiments contradict to our effect [3-7,9], is
incorrect.

\section*{Acknowledgements}

We are thankful to N.N. Bagmet, V.A. Belinski, Steven Carlip,
Douglas Singleton, Elias Vagenas, and V.E. Zakharov for useful
discussions.

$^*$Also at: L.D. Landau Institute for Theoretical Physics, 2
Kosygina Street, Moscow 117334, Russia.


\end{document}